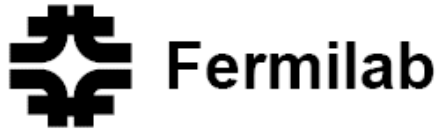

# RADIATION EFFECTS IN A MUON COLLIDER RING AND DIPOLE MAGNET PROTECTION*†

N.V. Mokhov#, V.V. Kashikhin, I. Novitski, A.V. Zlobin
Fermilab, Batavia, IL 60510, U.S.A.

## Abstract

The requirements and operating conditions for a Muon Collider Storage Ring (MCSR) pose significant challenges to superconducting magnets. The dipole magnets should provide a high magnetic field to reduce the ring circumference and thus maximize the number of muon collisions during their lifetime. One third of the beam energy is continuously deposited along the lattice by the decay electrons at the rate of 0.5 kW/m for a 1.5-TeV c.o.m. and a luminosity of $10^{34}$ cm$^{-2}$s$^{-1}$. Unlike dipoles in proton machines, the MCSR dipoles should allow this dynamic heat load to escape the magnet helium volume in the horizontal plane, predominantly towards the ring center. This paper presents the analysis and comparison of radiation effects in MCSR based on two dipole magnets designs. Tungsten masks in the interconnect regions are used in both cases to mitigate the unprecedented dynamic heat deposition and radiation in the magnet coils.

*Work supported by Fermi Research Alliance, LLC under contract No. DE-AC02-07CH11359 with the U.S. Department of Energy through the US LARP Program.
†Presented paper at Particle Accelerator Conference'11, March 28 – April 1, 2011, New York, U.S.A.
#mokhov@fnal.gov

# Radiation Effects in a Muon Collider Ring and Dipole Magnet Protection*

N.V. Mokhov[#], V.V. Kashikhin, I. Novitski, A.V. Zlobin
Fermilab, Batavia, IL 60510, U.S.A.

*Abstract*

The requirements and operating conditions for a Muon Collider Storage Ring (MCSR) pose significant challenges to superconducting magnets. The dipole magnets should provide a high magnetic field to reduce the ring circumference and thus maximize the number of muon collisions during their lifetime. One third of the beam energy is continuously deposited along the lattice by the decay electrons at the rate of 0.5 kW/m for a 1.5-TeV c.o.m. and a luminosity of $10^{34}$ cm$^{-2}$s$^{-1}$. Unlike dipoles in proton machines, the MCSR dipoles should allow this dynamic heat load to escape the magnet helium volume in the horizontal plane, predominantly towards the ring center. This paper presents the analysis and comparison of radiation effects in MCSR based on two dipole magnets designs. Tungsten masks in the interconnect regions are used in both cases to mitigate the unprecedented dynamic heat deposition and radiation in the magnet coils.

## INTRODUCTION

A number of demanding requirements to the collider optics and magnets result from the short muon lifetime, limitations on the dipole and quadrupole field quality and margins [1], and the necessity to protect superconducting magnets from muon decay products at the rate of ~0.5 kW/m for 750-GeV muon beams [2, 3].

Two alternative designs, one based on an open midplane approach with block type coils and absorber outside the coils, and the other based on a traditional large-aperture cos-theta approach with a shifted beam pipe and absorber inside the coil aperture were developed for the MCSR [2]. It was found that field quality and stress issues in the block-coil open midplane dipoles are quite severe and thus need more studies. Furthermore, MARS studies have shown (see below) that the position of the hottest spot in the dipole coils alternates along the ring lattice which makes the cos-theta coil design with closed midplane and asymmetric central absorber [4] less straightforward for use in the MCSR lattice. Therefore, in this paper we have focused on the open midplane dipole design with shell-type coils which provides good field quality in a reasonably large aperture.

While the detailed coil and support structure optimization is still a subject to a separate analysis, it was nevertheless demonstrated that the design with midplane spacers can structurally withstand the operating conditions.

## MCSR MAGNET DESIGN

The specifics of the heat deposition distributions in the MCSR dipoles – with decay products inducing showers predominantly in the orbit plane – require either a very large aperture with massive high-Z absorbers to protect the coils or an open midplane design. It has been shown [1-4] that the most promising approach is the open mid-plane design which allows decay electrons to pass between the superconducting coils and be absorbed in high-Z rods cooled at liquid nitrogen temperatures, placed far from the coils. MCR parameters used in this paper are given in Table 1.

Table 1: MC Storage Ring Parameters.

| Parameter | Unit | Value |
|---|---|---|
| Beam energy | TeV | 0.75 |
| Nominal dipole field | T | 10 |
| Circumference | km | 2.5 |
| Momentum acceptance | % | ±1.2 |
| Transverse emittance, $\varepsilon_N$ | $\pi$·mm·mrad | 25 |
| Number of IPs | | 2 |
| $\beta^*$ | cm | 1 |

The dipole coils used in this study are arranged in a shell-type configuration. The coil aperture is 80 mm, the coil to coil gap is 30 mm with supporting Al-spacers, and magnetic length is 6 m. The nominal field is 10 T. The relatively high level of magnetic fields in MCSR magnets needed for a reliable operation margin suggests using Nb$_3$Sn superconductor. The MCSR dipole coil cross-section, as produced by the ROXIE code [5], is shown in Fig. 1. The parameters for this and other possible magnet designs are reported in [1]. Although, the radiation analysis was performed for the shell-type open midplane magnet, it is also applied to the block type design.

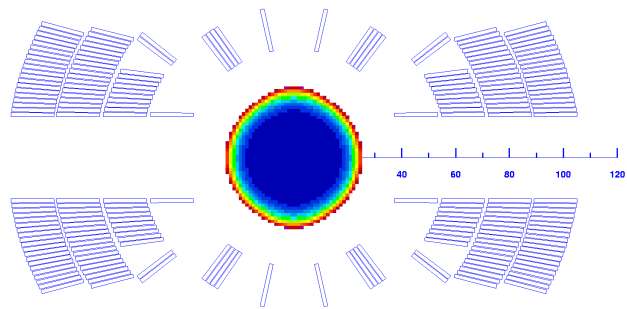

Figure 1. MCSR dipole based on 4-layer shell-type coil.

___________________
*Work supported by Fermi Research Alliance, LLC under Contract DE-AC02-07CH11359 with the U.S. DOE.
[#]mokhov@fnal.gov

# ENERGY DEPOSITION IN ARC MAGNETS

Energy deposition and detector backgrounds are simulated with the MARS15 code [6]. All the related details of geometry, materials distributions and magnetic fields are implemented into the model for lattice elements and tunnel in the interaction region and adjacent arcs, detector components, experimental hall and machine-detector interface. Fig. 2 shows the MARS model for the MCSR dipole. To protect the superconducting magnets and detector, 10 and 20-cm tungsten masks - with 5 $\sigma_{x,y}$ elliptic openings within 53 m from the IP and a larger round aperture at larger distances were implemented in the magnet interconnect regions in the model and carefully optimized. Two 750-GeV one-bunch muon beams – initially with $2\times10^{11}$ muons in each bunch - are assumed to be aborted after 1000 turns. The cut-off energy for all particles but neutrons is 200 keV, neutrons are followed down to the thermal energy (~0.001 eV)

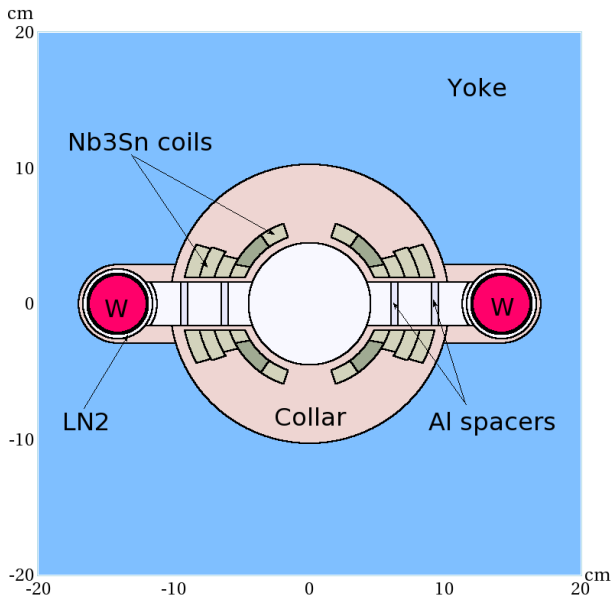

Figure 2: MARS model of the arc dipole magnet.

Power density profiles in mW/g (equal numerically to absorbed dose in Gy/s) in the orbit plane (±1.5-cm layer) are shown in Fig. 3 for the magnets of the first 100 m from the IP. The geometry and results are given in the beam coordinate system. With the masks in place and two circulating beams, the distributions along the arc magnet lengths are relatively uniform, although always with elevated levels at the magnet upstream and downstream ends.

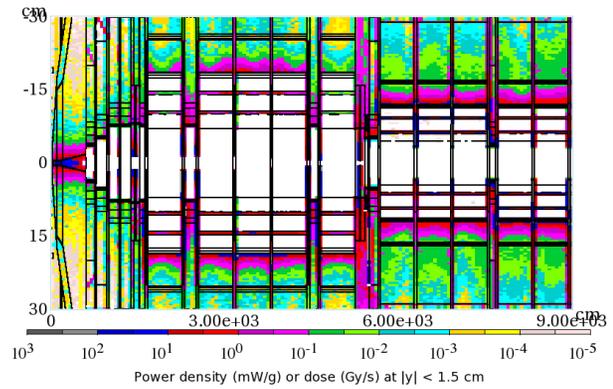

Figure 3: Power density distribution in magnets of the first 100 m from the IP (beam coordinate system).

Figs. 4 and 5 show power density distributions in two representative arc dipoles. The right side in these plots is toward the ring center. The open midplane design of the dipoles allows their safe operation. Four 7-mm wide aluminum spacers in the gap are found to have a minimal impact on the coil heating. The peak power density in the dipole inner coils ranges from 1.5 to 4 mW/g, safely below the quench limit for the $Nb_3Sn$ based superconducting coils in 1.9K operation.

The tungsten rods cooled by liquid nitrogen reduce the dynamic heat load on the liquid helium cryogenic system by almost a factor of two: 200 out of 445 W/m is dissipated in the rods in the dipole magnets shown in Fig. 4 and 5.

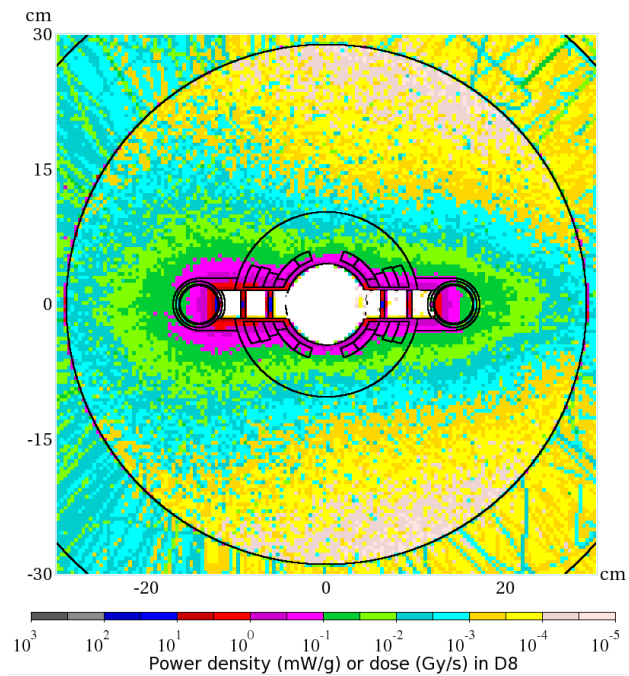

Figure 4: Power density (absorbed dose) profiles in a MCSR dipole 1.

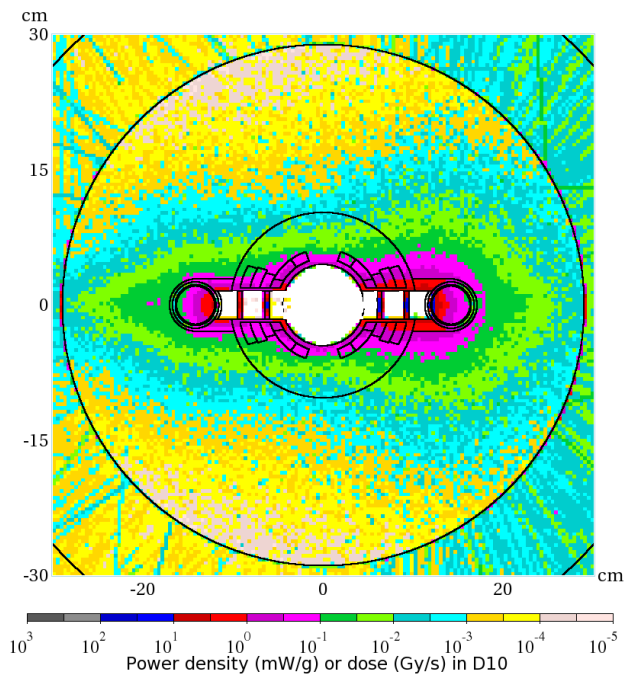

Figure 5: Power density (absorbed dose) profiles in a MCSR dipole 2.

The masks in the interconnect regions drastically reduce the heat load to the MCSR magnets. In the ring design with open midplane dipoles, energy deposition in the quadrupoles is very sensitive to the mask parameters. Radiation from a dipole midplane gap punches through a thin mask or outside it, and causes high local energy deposition in the quadrupole. Fig. 6 shows results for such a "minimal" mask case.

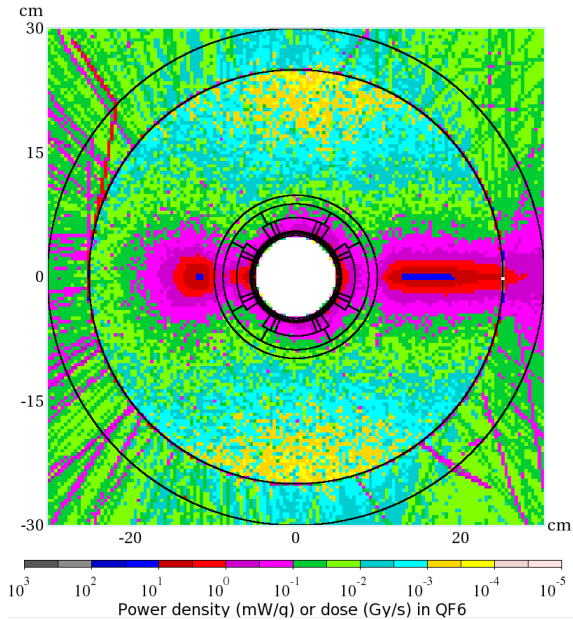

Figure 6: Power density (absorbed dose) profiles in a quadrupole magnet.

For the design studied here in detail, with 10-cm thick tungsten masks of a 7-cm round aperture and 20-cm outer diameter, the peak power density in the quadrupole coils ranges from a few mW/g to 30-40 mW/g, above the $Nb_3Sn$ quench limit. At the same time, a downstream dipole magnet is nicely protected by a combination of two masks and quadrupole material/field. Increasing the mask thickness to 20 cm and outer diameter to 30 cm, substantially reduces the heat load to the quadrupoles. Further optimization of the masks and consideration of high-Z liners in the quadrupole aperture is needed.

Radioactivation of the collider ring magnets is substantially lower than that in hadron colliders, because of the predominantly electromagnetic nature of the radiation in MCSR. Fig. 7 shows residual dose isocontours in one of the hot quadrupoles. The peak values on the cold mass outside reaches only 1 mSv/hr. In this respect a muon collider is a very clean machine.

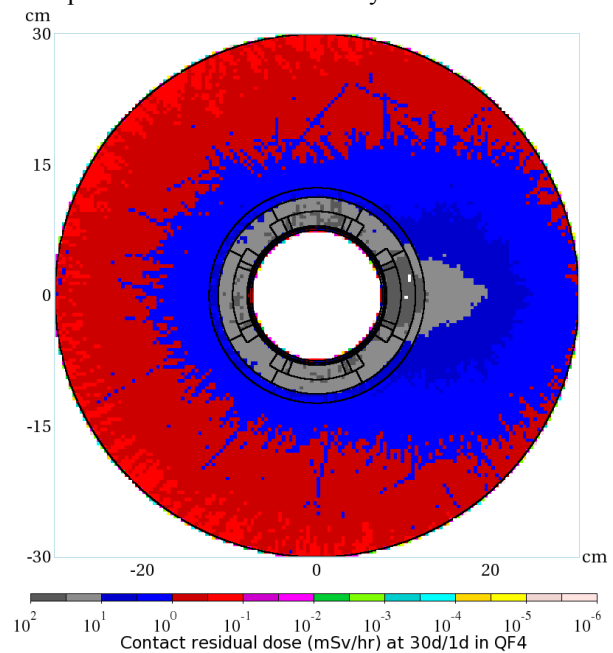

Figure 7: Residual dose profiles in a MCSR quadrupole.